\documentclass[aps,prl,reprint,citeautoscript, floatfix]{revtex4-1}

\usepackage{graphicx}
\usepackage{dcolumn}
\usepackage{bm}
\usepackage{braket}   
\usepackage[colorlinks, linkcolor=blue, citecolor=blue, urlcolor=blue]{hyper ref}
\usepackage{url}

\usepackage{natbib}
\usepackage{datatool}
\usepackage{etoolbox}

\DTLnewdb{bibnotes}

\begin{document}

\preprint{APS/123-QED}

\title{Trap-Assisted Auger-Meitner Recombination from First Principles}

\author{Fangzhou Zhao$^1$}
\author{Mark E. Turiansky$^1$}
\author{Audrius Alkauskas$^2$}
\author{Chris G. Van de Walle$^1$}
\email{vandewalle@mrl.ucsb.edu}
\affiliation{$^1$ Materials Department, University of California, Santa Barbara, CA 93106-5050, U.S.A.}
\affiliation{$^2$ Center for Physical Sciences and Technology (FTMC), Vilnius LT-10257, Lithuania}

\date{\today}

\begin{abstract}
Trap-assisted nonradiative recombination is known to limit the efficiency of optoelectronic devices, but the conventional multi-phonon emission (MPE) process fails to explain the observed loss 
in wide-band-gap materials.
Here we highlight the role of trap-assisted Auger-Meitner (TAAM) recombination, and present a first-principles methodology to determine TAAM rates due to defects or impurities in semiconductors or insulators. 
We assess the impact on efficiency of light emitters in a recombination cycle that may include both TAAM and carrier capture via MPE.
We apply the formalism to the technologically relevant case study of a calcium impurity in InGaN, where a Shockley-Read-Hall recombination cycle involving MPE alone cannot explain the experimentally observed nonradiative loss.
We find that, for band gaps larger than 2.5 eV, the inclusion of TAAM results in recombination rates that are orders of magnitude larger than recombination rates based on MPE alone, demonstrating that TAAM can be a dominant nonradiative process in wide-band-gap materials. 
Our computational formalism is general and can be applied to the calculation of TAAM rates in any semiconducting or insulating material. 

\end{abstract}

\maketitle

\begin{figure}[b]
\includegraphics[width=86mm]{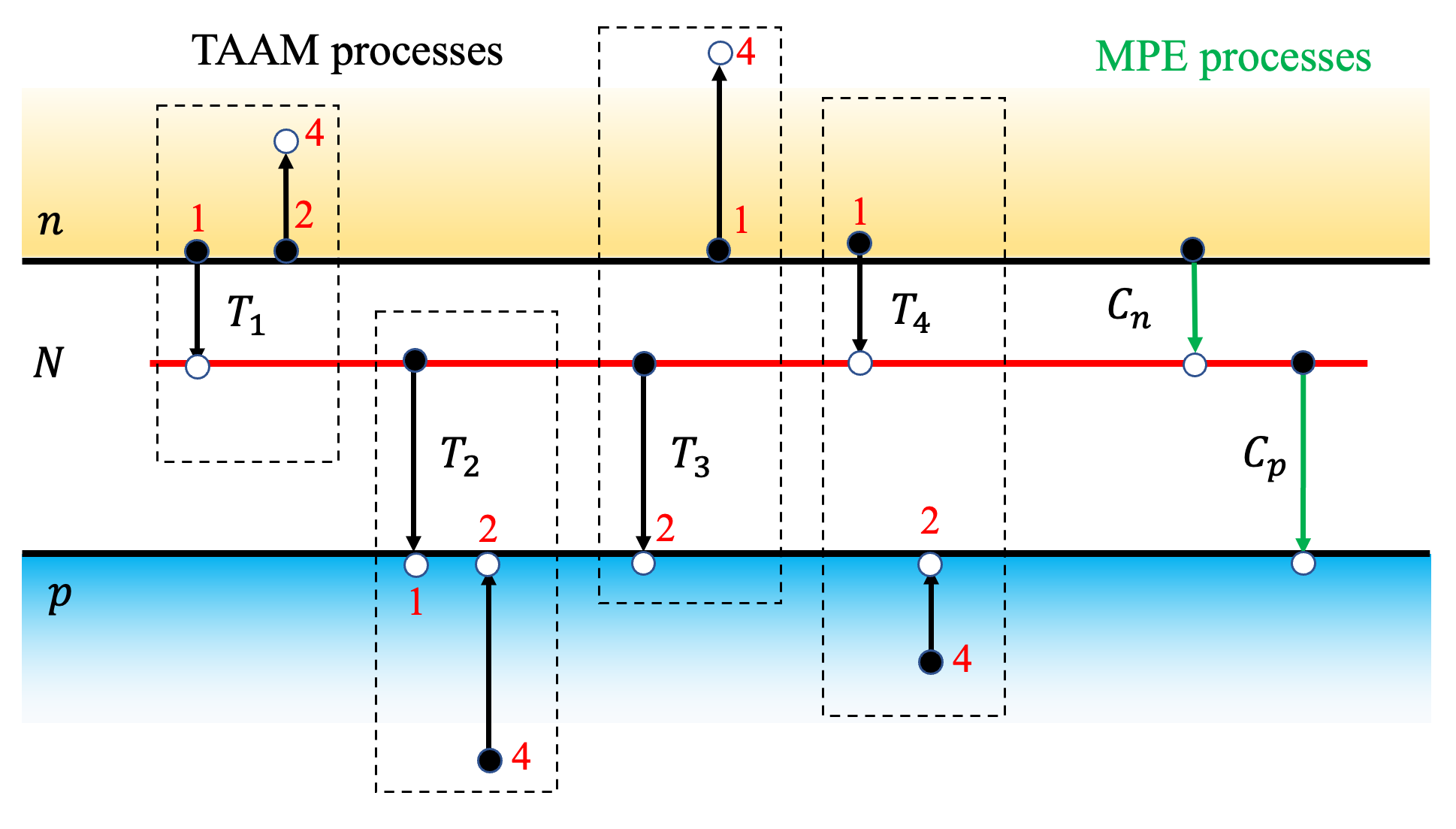}
\caption{\label{fig:TAAMschematic} Schematic diagram of the trap-assisted nonradiative recombination processes considered in this work: the four TAAM  processes and the two MPE processes. The TAAM processes, labeled by $T_1$, $T_2$, $T_3$, $T_4$, are depicted in the dashed rectangles. 
Solid (hollow) circles denote electrons (holes). The arrows denote the electron transitions, and the red numbers denote the state numbers in Eqs.~(\ref{eq:originalT1})-(\ref{eq:originalT4}). 
}
\end{figure}

Auger-Meitner recombination \cite{meitner1922entstehung,auger1923rayons,matsakis2019renaming} is an important nonradiative carrier recombination mechanism that has been widely invoked as a loss mechanism in optoelectronic devices \cite{schubert2006light}.
The Auger-Meitner process involves an electron-hole recombination event with the energy transferred via Coulomb interaction to a third free carrier that is excited to a higher-energy state. 
The bulk Auger-Meitner process based on free carriers in the conduction bands (CBs) valence band (VBs) scales as the third power of the carrier density; it therefore dominates at high carrier densities and has been identified as responsible for the efficiency droop of solid-state light emitters \cite{gardner2007blue, shen2007auger, kioupakis2011indirect, iveland2013direct, kioupakis2015first}.
In a trap-assisted Auger-Meitner (TAAM) process (Fig.~\ref{fig:TAAMschematic}), one of the carriers is localized on a point defect or impurity \cite{bess1958radiationless}, and hence the recombination rate scales as the second power of the carrier density \cite{landsberg1964auger, abakumov1991nonradiative}. 
This scaling allows the process to be distinguished from 
carrier capture by multiphonon emission (MPE) \cite{abakumov1991nonradiative}, which scales linearly with carrier density. 
TAAM recombination has occasionally been invoked as impacting the performance of semiconductor devices \cite{hangleiter1985experimental, cohn2013size, espenlaub2019evidence, myers2020evidence}; however, systematic studies are still lacking. 
Theoretical study has been based on analytic models \cite{haug1980auger} or focused on a specific scenario \cite{siyushev2013optically}.
Here, we present a general first-principles formulation along with a computationally feasible implementation and an assessment of the impact on efficiency-limiting nonradiative recombination. 

We illustrate the development of the formalism and the power of the approach by applying it to a relevant case study, namely a calcium impurity in InGaN, a key material for solid-state lighting.
Unintentionally incorporated calcium was experimentally observed to severely impact the quantum efficiency of light-emitting diodes \cite{young2016calcium}.
First-principles calculations of carrier capture via MPE \cite{shen2017calcium}, which is the most frequently discussed defect-assisted nonradiative recombination mechanism, indeed indicated that Ca acts as a strong Shockley-Read-Hall (SRH) \cite{shockley1952statistics, hall1952electron} recombination center in InGaN with a band gap up to $\sim$2.5 eV.
For larger band gaps, however, Ca-assisted SRH rates become vanishingly small, because the capture rate via MPE decreases exponentially as the energy difference between the trap level and the band edge increases.
Nonradiative SRH recombination based on MPE alone thus cannot explain the poor quantum efficiency in Ca-containing In$_{0.1}$Ga$_{0.9}$N layers with gaps close to 3 eV \cite{young2016calcium}.
More generally, this rapid decrease in capture rate for defect levels farther from the band edges, combined with the fact that carrier capture from both VB and CB is necessary for a complete nonradiative recombination cycle, means that capture via MPE cannot explain defect-assisted efficiency loss in wider-band-gap semiconductors.

In this Letter, we show that TAAM recombination provides a compelling explanation for nonradiative loss in wide-band-gap semiconductors. 
Four distinct TAAM processes can occur at a trap with a single bound state (Fig.~\ref{fig:TAAMschematic}), characterized by the coefficients $T_1$ through $T_4$ \cite{abakumov1991nonradiative}. 
For the capture of a free electron by the trap state, energy conservation is provided by exciting either a second free electron (process $T_1$) or a hole ($T_4$) to a higher-energy state.
Similarly, hole capture at the trap can be accompanied by the excitation of another hole ($T_2$) or an electron ($T_3$). 

Figure~\ref{fig:TAAMschematic} also depicts the single-carrier processes, in which the energy resulting from electron or hole capture is released via MPE.
Without loss of generality, we will consider the trap to be an acceptor, where the neutral charge state (with density $N^0$) is the initial state for electron capture, and the negative charge state (with density $N^-$) the initial state for hole capture.
The MPE rate for electron (hole) capture is linear in the carrier density $n$ ($p$) and given by $R_n=C_n N^0 n$ ($R_p=C_p N^- p$) \cite{alkauskas2014first}.
The electron (hole) capture coefficients $C_n$ ($C_p$) have units $\rm cm^3 s^{-1}$.
Recombination rates for the TAAM processes $R_{n,i}$ ($i=1,4$), $R_{p,i}$ ($i=2,3$) can similarly be expressed in terms of the coefficients $T_i$ where $n$ ($p$) denotes electron (hole) capture; these rates again scale with the trap density ($N^0$ or $N^-$) but they are {\em second order} in the carrier densities, since {\em two} free carriers are involved.
These rates are calculated based on Fermi’s golden rule and lead to the following expressions for the coefficients $T_i$ \cite{haug1980auger}:
\begin{widetext}
\begin{eqnarray}
  T_1 &&= \frac{ R_{n,1}}{N^0 n^2 } = \frac{2\pi}{ \hbar}  \frac{1}{ n^2 }  \sum_{\bm 1 \in c, \bm 2 \in c,  \bm 4 \in c}  f_{\bm 1} f_{\bm 2} (1-f_{\bm 4}) | M^1_{\bm 1\bm 2\bm t\bm 4} |^2  \delta (\epsilon_{\bm 1}+ \epsilon_{\bm 2} - \epsilon_{\bm t} - \epsilon_{\bm 4})  , \label{eq:originalT1} \\
  T_2 &&= \frac{R_{p,2}}{N^- p^2 } = \frac{2\pi}{ \hbar}  \frac{1}{ p^2 }  \sum_{\bm 1 \in v, \bm 2 \in v,  \bm 4 \in v} (1- f_{\bm 1}) (1 - f_{\bm 2})  f_{\bm 4} | M^2_{\bm 1\bm 2\bm t\bm 4} |^2  \delta (\epsilon_{\bm 1}+ \epsilon_{\bm 2} - \epsilon_{\bm t} - \epsilon_{\bm 4}) , \label{eq:originalT2}\\
  T_3 &&= \frac{R_{p,3}}{N^- n p } = \frac{2\pi}{ \hbar}  \frac{1}{ np }  \sum_{\bm 1 \in c, \bm 2 \in v,  \bm 4 \in c}  f_{\bm 1} (1- f_{\bm 2}) (1-f_{\bm 4}) | M^3_{\bm 1\bm 2\bm t\bm 4} |^2  \delta (\epsilon_{\bm 1}+ \epsilon_{\bm t} - \epsilon_{\bm 2} - \epsilon_{\bm 4}) , \label{eq:originalT3}\\
  T_4 &&= \frac{R_{n,4}}{N^0 n p } = \frac{2\pi}{ \hbar}  \frac{1}{ np }  \sum_{\bm 1 \in c, \bm 2 \in v,  \bm 4 \in v}  f_{\bm 1} (1- f_{\bm 2} ) f_{\bm 4} | M^4_{\bm 1\bm 2\bm t\bm 4} |^2  \delta (\epsilon_{\bm 1}+ \epsilon_{\bm 4} - \epsilon_{\bm t} - \epsilon_{\bm 2}) ,  \label{eq:originalT4}
\end{eqnarray}
\end{widetext}
where $c$ and $v$ indicate the bulk states in the CB and VB continuum, and the trap-state level is labeled by $t$. 
$f_{\bm j}$ are free-carrier occupation numbers for the $\bm j$-th carrier state according to Fermi-Dirac statistics. The $\delta$ function ensures energy conservation. The units of $R_{n/p,i}$ are $\rm cm^{-3} s^{-1}$, and the units of $T_i$ are $\rm  cm^6 s^{-1}$. 
Vibrational broadening is included by replacing 
Eqs.~(1)-(4) by a convolution with a normalized spectral function of electron-phonon interaction \cite{alkauskas2012first, razinkovas2021photoionization}.

The matrix elements $M_{\bm 1\bm 2\bm t\bm 4}^i$ for the four processes are given by:
\begin{eqnarray}
 M^1_{\bm 1\bm 2\bm t\bm 4} && = M^2_{\bm 1\bm 2\bm t\bm 4} =\bra{\bm 1\bm  2} \hat{W} \ket{\bm t \bm 4} - \bra{\bm 1\bm  2} \hat{W} \ket{\bm 4 \bm t}, \\
 M^3_{\bm 1\bm 2\bm t\bm 4} && = \bra{\bm 1 \bm t} \hat{W} \ket{\bm  4 \bm 2} - \bra{\bm 1 \bm t} \hat{W} \ket{\bm 2 \bm 4}, \\
 M^4_{\bm 1\bm 2\bm t\bm 4} && = \bra{\bm 1 \bm 4} \hat{W} \ket{\bm t \bm 2} \bra{\bm 1 \bm 4} \hat{W} \ket{\bm 2 \bm t},
\end{eqnarray}
where each number or $\bm t$ indicates the band indices and spin indices for a spin-polarized calculation, or the spinor wavefunction states in a noncollinear calculation. 
The matrix elements of the screened Coulomb interaction $\hat{W}$ are given by 
\begin{eqnarray}
 && \bra{\bm 1 \bm 2} \hat{W} \ket{\bm t \bm 4} = \nonumber  \\
 && \int \int d \bm r_1 d \bm r_2 \psi_{\bm 1}^* (\bm r_1 ) \psi_{\bm 2}^* (\bm r_2 )W(\bm r_1,\bm r_2 ) \psi_{\bm t} (\bm r_1 ) \psi_{\bm 4} ( \bm r_2  )
\end{eqnarray}
with $\hat{W}$ approximated using a model dielectric function \cite{cappellini1993model} [see Supplemental Material (SM), S1 \cite{supplementary}]. 

As already mentioned, a complete SRH recombination cycle \cite{SRHnote} requires capture of both an electron and a hole.
Here we address this complete cycle by taking into account that capture could occur through either MPE or TAAM.
Assuming that $n=p$ (as is usually the case in light emitters due to charge neutrality), the total trap-assisted nonradiative recombination rate is then given by (see SM, S2 \cite{supplementary}): 
 \begin{equation}
 R_{tot} = N \frac{(T_1+T_4+C_n/n)(T_2+T_3+C_p/n)} {T_1+T_2+T_3+T_4+(C_n+C_p)/n} n^2
 \label{eq:Rtot}
 \end{equation}
where $N=N^0 + N^-$ is the total density of traps. 
We note that this rate is second order in $n$ at high carrier density, but still linear (as in the “usual” SRH recombination cycle) if the TAAM coefficients are small and the carrier density is low.

Our quantitative calculations of TAAM coefficients are based on first-principles density functional theory.  
To obtain accurate results for defects and impurities \cite{freysoldt2014first} we use the hybrid functional of Heyd-Scuseria-Ernzerhof (HSE) \cite{heyd2003hybrid, *heyd2006hybrid}. 
For our case study of Ca in GaN we use a 400-eV energy cutoff, and the fraction of screened Fock exchange $\alpha$ in the HSE functional is set to 0.31, which results in a GaN band gap of 3.55 eV, in agreement with experiment \cite{madelung2004semiconductors}. 
The Ga $d$ states are treated as part of the core and spin polarization is included.
Energetics are calculated in a 96-atom supercell with a $2\times 2 \times 2$ Monkhorst-Pack k-point grid, using the Vienna Ab-initio Simulation Package (VASP) \cite{kresse1996efficient, kresse1996efficiency} with projector-augmented wave (PAW) potentials. 
Finite-size corrections for charged systems are applied \cite{freysoldt2009fully, freysoldt2011electrostatic}. 
We evaluate 
the TAAM coefficients using the eigenvalues and wavefunctions from {\sc Quantum Espresso} (QE) \cite{giannozzi2009quantum} with norm-conserving pseudopotentials to avoid the complications in the matrix elements associated with the description of the PAW core region.
We used the $\Gamma$ point in 96-, 360-, and 768-atom supercells (see SM, S3 \cite{supplementary}), with structural relaxations performed within QE. A choice of $\alpha$=0.45 renders the QE results consistent with VASP, with $<50$ meV difference in energy levels and unobservable differences in wavefunctions. 

\begin{figure}[b]
\includegraphics[width=86mm]{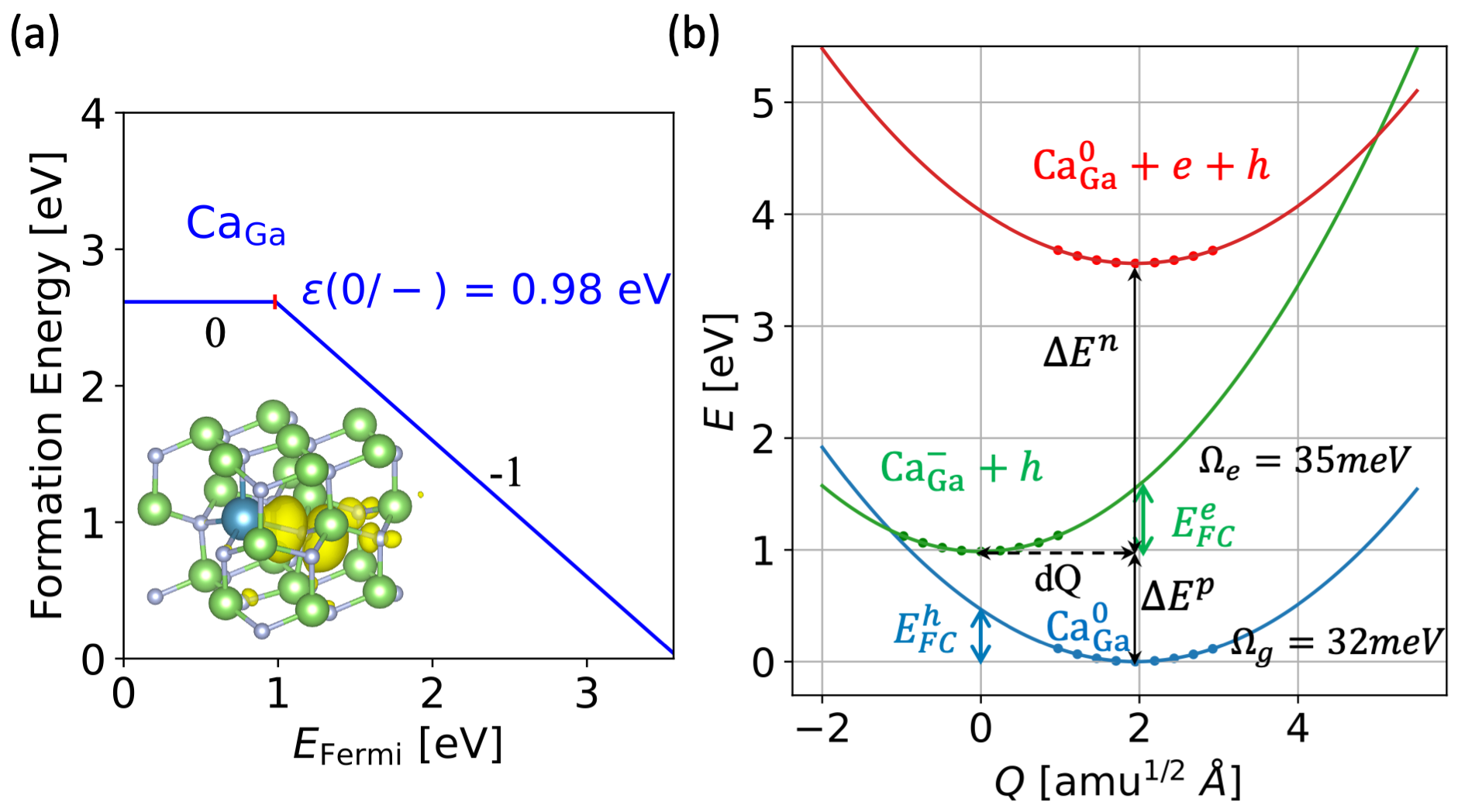}
\caption{\label{fig:Ca} (a) Formation energy vs. Fermi level for $\rm Ca_{Ga}$ in neutral and negative charge states under Ga-rich conditions. The atomic geometry and trap-state wavefunction of neutral $\rm Ca_{Ga}$ are illustrated in the inset. (b) Configuration coordinate diagram illustrating electron and hole capture processes.  The symbols denote calculated values; the lines are parabolic fits.  }
\end{figure}

Figure~\ref{fig:Ca}(a) shows the calculated formation energy of the substitutional $\rm Ca_{Ga}$ impurity, which was found to be relevant for nonradiative recombination \cite{shen2017calcium}.
$\rm Ca_{Ga}$ acts as a deep acceptor with a (0/$-$) level $\sim 0.98$ eV above the valence-band maximum (VBM). The inset in Fig.~\ref{fig:Ca}(a) shows the Kohn-Sham wavefunction of the single trap state in the gap in the minority-spin channel: it is a $p$-like orbital localized at a nitrogen atom adjacent to the Ca impurity atom.

Electron-phonon interactions, which we need for evaluation of the MPE \cite{turiansky2021nonrad} as well as to include vibrational broadening \cite{stoneham1981non} in the TAAM processes are evaluated based on a one-dimensional configuration coordinate diagram [Fig.~\ref{fig:Ca}(b)] \cite{alkauskas2016tutorial}, which is justified in the case of strong electron-phonon coupling \cite{alkauskas2012first}. 
The transition energy for electron capture $\Delta E^n$, also known as the zero-phonon line, is the energy difference between the conduction-band minimum (CBM) and the (0/$-$) transition level (and {\em mutatis mutandis} for hole capture $\Delta E^p$). 
The Franck-Condon energies are $E_{FC}^e$=0.54 eV for electron capture and $E_{FC}^h$=0.50 eV for hole capture, and the Huang-Rhys factors are $S^e=\frac{E_{FC}^e} {\hbar \Omega_e } \approx 15 $  and $S^h=\frac{E_{FC}^h} {\hbar \Omega_g} \approx 16$. 

In the case of strong electron-phonon coupling (large Huang-Rhys factors), the electron-phonon spectral function is well approximated by a Gaussian with a variance $\sigma$ 
determined based on the $S$ and $\Omega$ parameters \cite{jia2020design, alkauskas2016tutorial} (See SM, S5 \cite{supplementary}).
We include phonon assistance in the TAAM coefficient for the $T_1$ process by replacing  $\delta (\epsilon_{\bm 1} + \epsilon_{\bm 2} - \epsilon_{\bm t} - \epsilon_{\bm 4}) $ by a Gaussian function $g$ centered at [$ \epsilon_{\bm 1} + \epsilon_{\bm 2} - \epsilon (0/-) - \epsilon_{\bm 4} - E_{FC}^e$].
Similar expressions apply to the other processes. The energy of the KS state $\epsilon_t$ is replaced by [$\epsilon(0/-) + E^e_{FC}$] to reflect that the vertical transition energy [Fig.~\ref{fig:Ca}(b)] released by the first carrier is transferred to the second carrier. $\sigma$ is found to be close to 0.22 eV for all processes. 

\begin{figure}[b]
\includegraphics[width=86mm]{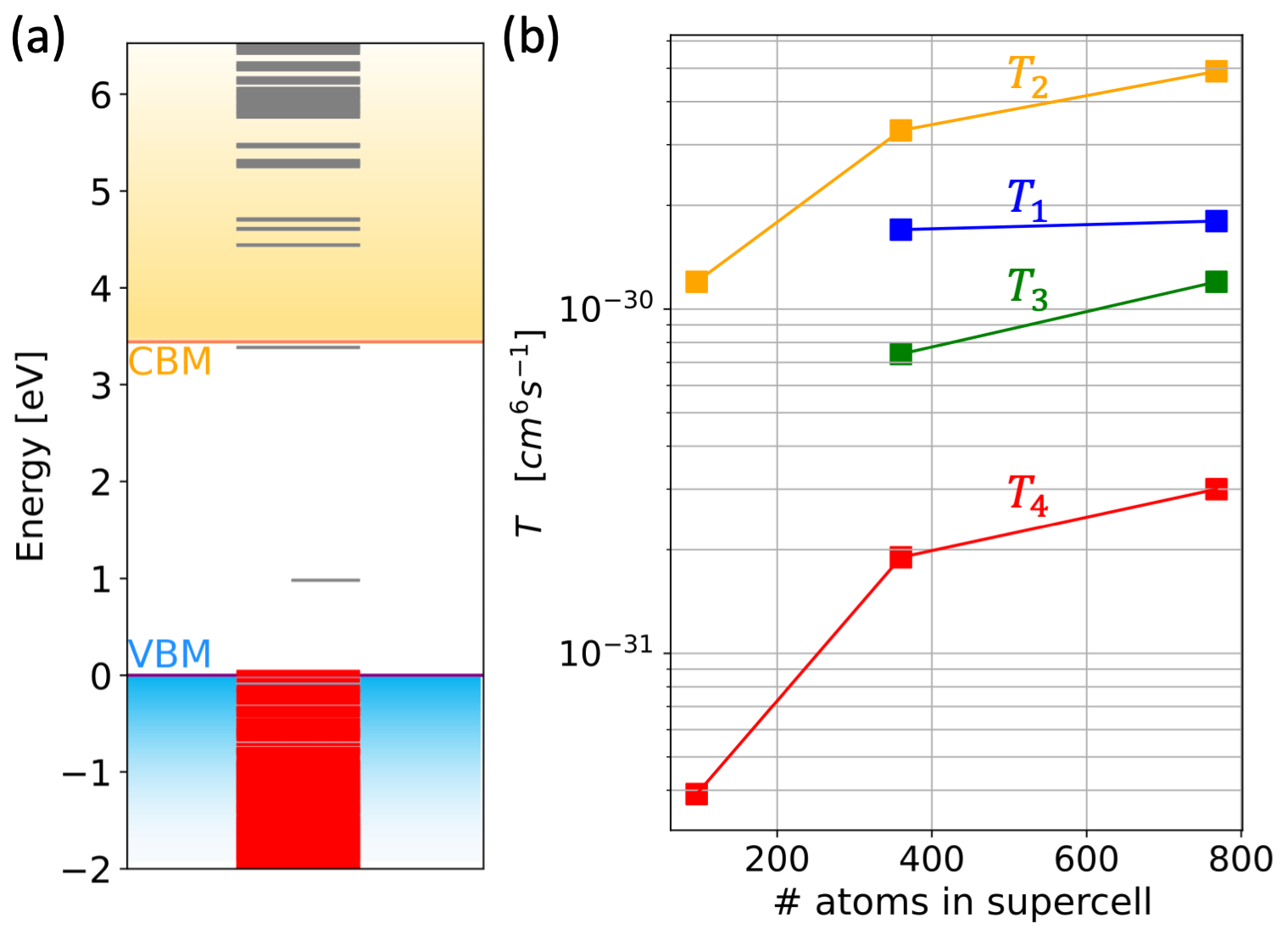}
\caption{\label{fig:sampling} 
(a) Kohn-Sham states for $\rm Ca_{Ga}$ in GaN calculated in a 768-atom supercell. 
(b) Calculated TAAM coefficients as a function of supercell size. 
 }
\end{figure}

Particular attention needs to be paid to the summation over the $4^{th}$ bulk state in Eqs.~(1)-(4).
This state is sparsely sampled in our supercell calculations
in cases $T_1$ and $T_3$ due to the highly dispersive nature of the lowest CB [as shown by sampled Kohn-Sham states in the CB continuum illustrated in Fig.~\ref{fig:sampling}(a)]. 
Sampling of the VB states (cases $T_2$ and $T_4$) is far easier thanks to lower dispersion and zone folding in the supercell.
For $T_1$ and $T_3$, we therefore perform the summation over the $4^{th}$ CB state in a more physical way by a continuous integration according to the bulk CB density of states $D(\epsilon)$:
 \begin{eqnarray}
T_1 &&= \frac{2\pi}{ \hbar}  \frac{1}{ n^2 }  \sum_{\bm 1 \in c, \bm 2 \in c} \int_{CBM}^{\infty}  d\epsilon_{\bm 4} D(\epsilon_{\bm 4}) f_{\bm 1} f_{\bm 2} (1-f_{\bm 4} (\epsilon_{\bm 4}))     \nonumber   \\
 && \times \overline{| M^1_{\bm 1\bm 2\bm t\bm 4} |^2}  \, g [\epsilon_{\bm 1}+ \epsilon_{\bm 2} - \epsilon (0/-)- \epsilon_{\bm 4} - E_{FC}^e]  \, .
 \label{eq:T1}
 \end{eqnarray}
The use of an average value $\overline{| M^1_{\bm 1\bm 2\bm t\bm 4} |^2}$ is justified because the orbital character of the CB, and hence the matrix element, varies little over the relevant energy range (see SM, S4 \cite{supplementary}).
The upper bound of the integration is chosen to fully include the (vibrationally broadened) energy-conserving transition. 
When calculating TAAM coefficients for a material with a highly dispersive VB the same technique should be applied to $T_2$ and $T_4$. 

Figure~\ref{fig:sampling}(b) shows our calculated values for the $T_i$ coefficients using 96-, 360-, and 768-atom supercells at $T$=390 K.
Values for $T_1$ and $T_3$ in the 96-atom supercell are not included due to the sparse sampling of CB states. 
Comparing the 360- and 768- atom supercell calculations, 
the coefficients are converged to within a factor of two,  a satisfactory level of accuracy.
Extensive checks indicated that the $T_i$ coefficients depend only weakly on the position of the trap-state level in the band gap or on the value of the broadening parameter (See SM, S5 \cite{supplementary}). 

We now investigate the TAAM coefficients and the total nonradiative recombination rate in InGaN alloys.
Since explicit alloy calculations are prohibitively expensive, we use interpolation procedures similar to previous work on bulk Auger-Meitner \cite{kioupakis2011indirect, kioupakis2015first} and SRH recombination \cite{dreyer2016gallium,shen2017calcium}.
VBM and CBM positions in $\rm In_xGa_{1-x}N$ for $x<0.5$ are taken from 
Ref.~\onlinecite{moses2010band}, and 
the (0/$-$) transition level is interpolated based on explicit calculations at discrete values of $x$ \cite{shen2017calcium} following the procedure outlined in Ref.~\onlinecite{dreyer2016gallium}. 
$T_i$ coefficients as a function of $\rm In_xGa_{1-x}N$ gap are then calculated using the 360-atom supercell based on wavefunctions for $\rm Ca_{Ga}$ in GaN  but with the band edges and trap-state level rigidly shifted as specified above.

\begin{figure}[b]
\includegraphics[width=86mm]{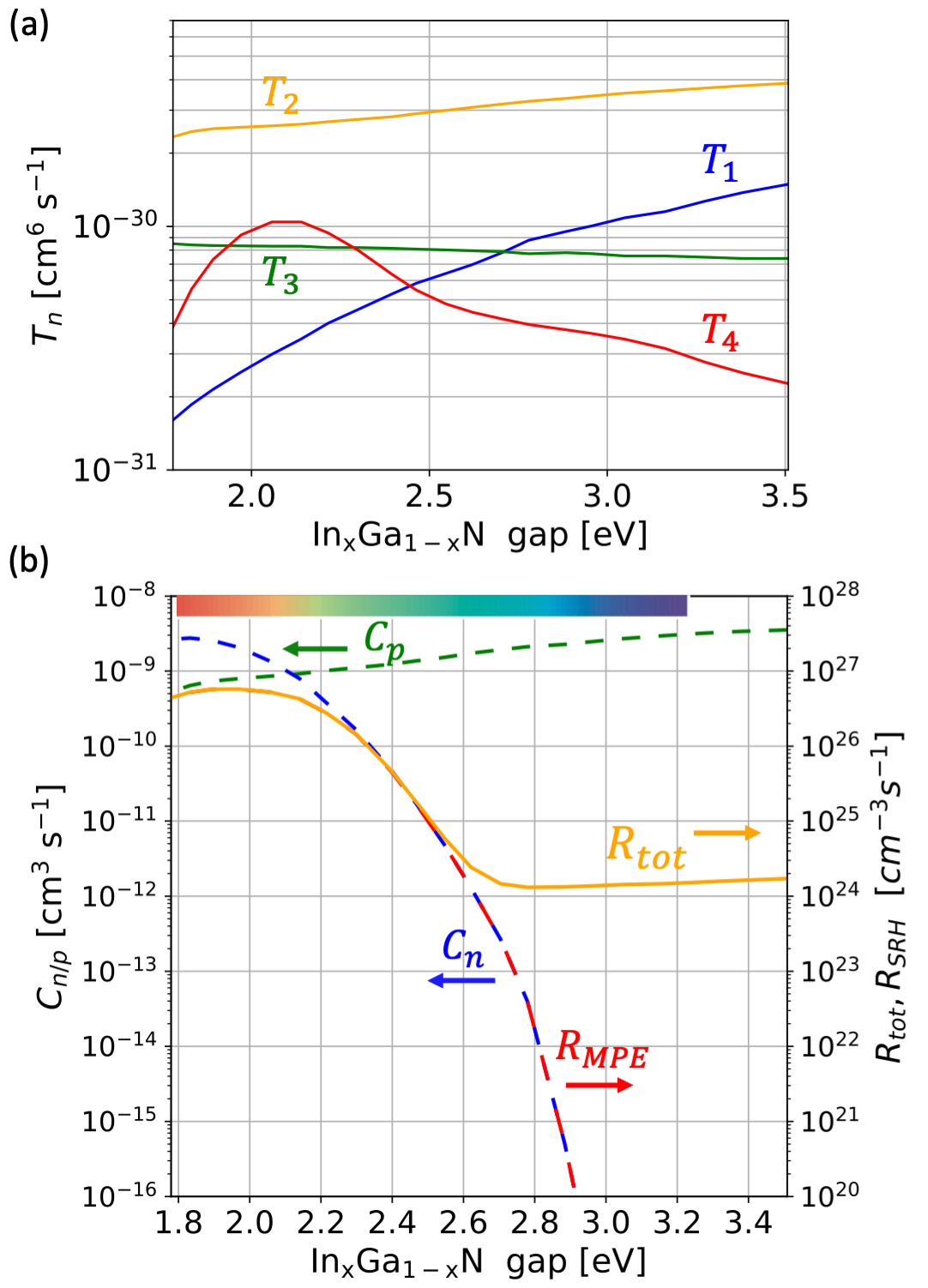}
\caption{\label{fig:InGaN} 
(a) Calculated TAAM coefficients for $\rm Ca_{Ga}$ as a function of $\rm In_xGa_{1-x}N$ band gap. 
(b) Calculated MPE capture coefficients $C_n$ and $C_p$ (dashed lines, left axis), as well as the total trap-assisted nonradiative recombination rate (right axis) calculated at $T$=390 K with $N = 10^{18}$ cm$^{-3}$ and $n = 10^{18}$ cm$^{-3}$.  $R_{tot}$ (orange curve) includes the two MPE processes plus the four TAAM processes [Eq.~(\ref{eq:Rtot})], while $R_{MPE}$ (red dashed curve) includes only the MPE processes.}
\end{figure}

Figure~\ref{fig:InGaN}(a) shows that all of the $T_i$ coefficients are on the order of $\rm 10^{-30}~cm^6 s^{-1}$.
The $T_2$ and $T_3$ coefficients are almost independent of $\rm In_xGa_{1-x}N$ gap since they are based on hole capture to the trap-state level, and the (0/$-$) level largely tracks the VBM \cite{shen2017calcium}. 
In contrast, $T_1$ and $T_4$ show larger variations, since the position of the (0/$-$) level relative to the CBM changes from 2.5 eV in GaN to 0.7 eV in In$_{0.5}$Ga$_{0.5}$N \cite{shen2017calcium}.
For $T_1$, a larger energy difference allows electron excitations to higher-lying CB states (Fig.~\ref{fig:TAAMschematic}) where the density of CB states is larger.
The variation trend is different for $T_4$, which involves excitation of a hole into the VB (see SM, S5 \cite{supplementary}).

Still, the variation of the $T_1$ and $T_4$ coefficients with band gap is relatively minor when compared to the huge change in the MPE capture coefficient $C_n$ [Fig.~\ref{fig:InGaN}(b)] \cite{MPE_calc_note}.
This is because, in a semiclassical picture, MPE capture depends exponentially on a barrier height that increases linearly with the energy difference between the trap level and the CBM \cite{abakumov1991nonradiative, alkauskas2016role}; no such activated behavior occurs in TAAM, where final states in the continuum are readily available at any energy.
The absence of activated behavior, and of the need for momentum conservation, also explain the TAAM coefficients' very weak dependence on temperature, as we have verified by explicit calculations.  This implies that TAAM recombination will persist as a loss mechanism even at low temperatures. \cite{hangleiter1987nonradiative}

These results now allow us to determine the total trap-assisted nonradiative recombination rate due to $\rm Ca_{Ga}$ in InGaN as a function of InGaN band gap. 
We use the $N=10^{18}$ cm$^{-3}$ Ca concentration from the experimental study \cite{young2016calcium} and a typical operating carrier density of $n=10^{18}$ cm$^{-3}$. 
Figure~\ref{fig:InGaN} compares the total recombination rate $R_{tot}$ with the rate $R_{MPE}$ assuming only MPE processes [obtained from Eq.~(\ref{eq:Rtot}) by setting all $T_i$ to zero]. 
For band gaps less than  $\sim$2.5 eV the rate is dominated by MPE capture; however, due to the increasing energy difference with the CBM, the MPE-assisted electron capture rate rapidly decreases with band gap. 
Around 2.5 eV TAAM becomes the dominant electron capture process, and due to its relative insensitivity to band gap [Fig.~\ref{fig:InGaN}(a)] the overall $R_{tot}$ remains relatively constant (or even slightly increases) as a function of band gap.

For In$_{0.1}$Ga$_{0.9}$N (as used in Ref.~\cite{young2016calcium}) the gap is $\sim$3.0 eV and the calculated $R_{tot}$ is $2 \times 10^{24} ~\rm cm^{-3} s^{-1} $, 11 orders of magnitude larger than the rate based on the MPE process alone. 
To put this in perspective, the calculated radiative recombination rate for In$_{0.1}$Ga$_{0.9}$N is about $4 \times 10^{25} \rm cm^{-3} s^{-1}$ \cite{kioupakis2013temperature}, 
so, to within the calculation error bars, nonradiative recombination due to Ca can significantly impact efficiency, as experimentally observed \cite{young2016calcium}---but it is essential to include TAAM processes.

Based on our calculated numbers, at band gaps where $C_n << n(T_1 + T_4)$ the expression for $R_{tot}$ can be approximated as $R_{tot} \approx N (T_1 + T_4) n^2$.
Espenlaub {\it et al.} \cite{espenlaub2019evidence} indeed observed the presence of a nonradiative recombination mechanism scaling as $n^2$; and Myers {\it et al.} \cite{myers2020evidence} found evidence for hot electrons in the conduction band, consistent with our calculations that show the $T_1$ electron capture process to be dominant.

Unlike MPE rates, which decrease exponentially with band gap  \cite{abakumov1991nonradiative, alkauskas2016role}, TAAM processes are not suppressed in large-gap materials because they are based on Coulomb interactions that can occur at any energy and for which final states are always available. 
As a result, inclusion of the TAAM-assisted processes can account for the observed nonradiative recombination rate in wider-band-gap materials, where rates due to MPE alone become negligibly low.

In conclusion, we have developed a first-principles formalism to calculate trap-assisted Auger-Meitner recombination rates. 
For our test case of $\rm Ca_{Ga}$ impurities in InGaN, the results show that including TAAM processes is essential to explain the observed trap-assisted nonradiative recombination rates in materials with band gaps greater than $\sim$ 2.5 eV. 
Our formalism is general and can be applied to study TAAM recombination in any semiconductor or insulator.  
The approach provides insight into the physics of nonradiative recombination processes and elucidates why TAAM processes are key to describing defect-assisted recombination in wider-band-gap materials, where MPE alone fail to explain efficiency loss.

\begin{acknowledgments}
We acknowledge fruitful discussions with E. Kioupakis. 
This study was supported by the US Department of Energy (DOE), Office of Science, Basic Energy Sciences (BES) under Award No. DE-SC0010689. 
F.Z. acknowledges support from the California NanoSystems Institute for an Elings Prize Fellowship. 
M.E.T. was supported by the National Science Foundation (NSF) through Enabling Quantum Leap: Convergent Accelerated Discovery Foundries for Quantum Materials Science, Engineering and Information (Q-AMASE-i) Award No. DMR-1906325.
Computational resources were provided by the National Energy Research Scientific Computing Center, a DOE Office of Science User Facility supported by the Office of Science of the DOE under Contract No. DE-AC02-05CH11231. 

\end{acknowledgments}

\appendix

\nocite{manchon1970optical, barker1973infrared}
\bibliography{bibliography}
\end{document}


\preprint{APS/123-QED}

\title{Supplemental Material for \\
“Trap-Assisted Auger-Meitner Recombination from First Principles”}

\author{Fangzhou Zhao$^1$}
\author{Mark E. Turiansky$^1$}
\author{Audrius Alkauskas$^2$}
\author{Chris G. Van de Walle$^1$}
\email{vandewalle@mrl.ucsb.edu}
\affiliation{$^1$ Materials Department, University of California, Santa Barbara, CA 93106-5050, U.S.A.}
\affiliation{$^2$ Center for Physical Sciences and Technology (FTMC), Vilnius LT-10257, Lithuania}

\date{\today}

\maketitle

\section{S1. Screened Coulomb interaction matrix elements in trap-assisted Auger coefficient calculations }

The matrix elements of the screened Coulomb interaction in our trap-assisted Auger-Meitner (TAAM) coefficients calculations are given by: 
 \begin{eqnarray}
   \bra{\bm 1 \bm 2} \hat{W} \ket{\bm t \bm 4} = \int \int d \bm r_1 d \bm r_2 \psi_{\bm 1}^* (\bm r_1)  \psi_{\bm 2}^* (\bm r_2) W(\bm r_1, \bm r_2)  \psi_{\bm t} (\bm r_1)   \psi_{\bm 4} (\bm r_2)  \, .
   \label{eq:ME1}
 \end{eqnarray}
Our supercell calculations for the trap of interest are performed at the $\Gamma$ point, so the wavefunction of the bulk states (trap states) is simply $\psi_{\bm i} (\bm r)=u_{\bm i}(\bm r)$ ($\psi_{\bm t} (\bm r) = u_{\bm t} (\bm r)$), where $u_{\bm i} (\bm r)$ ($u_{\bm t} (\bm r)$) is the lattice periodic part of the Bloch states.  The Fourier transform of the Coulomb interaction $\hat{W}$ is given by:
 \begin{eqnarray}
 W(\bm r_1, \bm r_2 )= W(\bm r_1- \bm r_2 )= \frac{1}{V_{cell}}  \sum_{\bm q}  \frac{1}{\epsilon (\bm q)}   \frac{4\pi e^2} {q^2+\lambda^2 }  e^{i \bm q \cdot (\bm r_1- \bm r_2)} = \frac{1}{V_{cell}}  \sum_{\bm G}  \frac{1}{\epsilon (\bm G)}   \frac{4\pi e^2} {G^2+\lambda^2 }  e^{i \bm G \cdot (\bm r_1- \bm r_2)}  
 \end{eqnarray}
where $\bm G$ is the G vector corresponding to the supercell, since the Brillouin zone is sampled at the $\Gamma$ point. As a result, the matrix elements can be evaluated by two Fourier transforms in the code: 
 \begin{eqnarray}
    \label{eq:ME2}
\bra{\bm 1 \bm 2} \hat{W} \ket{\bm t \bm 4}  = \frac{1}{V_{cell}}  \sum_{\bm G} \int  d \bm r_1 u_{\bm 1}^*(\bm r_1) u_{\bm t} (\bm r_1) e^{i\bm G \cdot \bm r_1}  \frac{1}{\epsilon (\bm G)}   \frac{4\pi e^2} {G^2+\lambda^2 } \int  d \bm r_2 u_{\bm 2}^*(\bm r_2) u_{\bm 4} (\bm r_2)  e^{ - i\bm G \cdot \bm r_2} \, ,
 \end{eqnarray}
where $ \int  d \bm r_1 u_{\bm i}^*(\bm r_1) u_{\bm j} (\bm r_1) e^{i\bm G \cdot \bm r_1} = \sum_{\bm G_1} c_i^* (\bm G_1 ) c_j (\bm G_1- \bm G)$ and $c$ are the plane-wave components of the lattice-periodic part of the Bloch functions $u$. 

Consistent with first-principles calculations of bulk Auger-Meitner recombination \cite{kioupakis2015first} we calculated the matrix elements of the screened Coulomb interaction using a model dielectric function \cite{cappellini1993model}:
 \begin{eqnarray}
 \epsilon(\bm G)=1+ \left[(\epsilon_{\infty}-1)^{-1} + \alpha ( \frac{q}{q_{TF}} )^2+  \frac{\hbar^2 q^4}{4m^2 \omega_p^2} ) \right]^{-1} \, ,
 \end{eqnarray}
where $\epsilon_{\infty}$ is the dielectric constant due to electronic screening, $q_{TF}$ is the Thomas-Fermi wave vector, $\omega_p$ is the plasma frequency, and $\alpha = 1.563$ is an empirical parameter. For the dielectric constant of GaN, we used the directionally averaged experimental dielectric constant, $\epsilon_{\infty}= (\epsilon_{\infty \parallel} \epsilon_{\infty \perp}^2 )^{1/3}=5.5 $ \cite{manchon1970optical, barker1973infrared} .  For the free-carrier screening wave vector $\lambda$, the Debye-Hückel equation for nondegenerate carriers $\lambda^2=\frac{4\pi ne^2}{\epsilon_{\infty} k_B T} $ is used. 

\section{S2. Total trap-assisted nonradiative recombination rate including trap-assisted Auger-Meitner and multiphonon emission processes}

In this section, we derive the formula for the total trap-assisted nonradiative recombination rate including the four TAAM and two multiphonon emission (MPE) carrier capture processes. As shown in Fig.~1 in the main text, the TAAM processes $T_1$ and $T_4$ capture an electron, and $T_2$ and $T_3$ capture a  hole. Without loss of generality, an acceptor trap state is considered, with a density $N^-$ for the negatively charged acceptor and a density $N^0$ for the neutral acceptor. 
The rate equations for the electron ($n$) and hole ($p$) carrier densities are then given by:
\begin{eqnarray}
\frac{dn}{dt}&&=-T_1 N^0 n^2 - T_4 N^0 np - C_n N^0 n + G                  \\
\frac{dp}{dt}&&=-T_2 N^- p^2 - T_3 N^- np - C_p N^- p + G
 \end{eqnarray}
where $G$ is the carrier generation rate.
In steady state, $\frac{dn}{dt} = \frac{dp}{dt} =0 $, the carrier concentrations and capture rates stay constant. 
Using $N = N^- + N^0$ we can then solve for the occupation of the trap state:
 \begin{eqnarray}
\frac{N^-}{N} = \frac{T_1 n^2+T_4 np+C_n n}  {T_1 n^2+T_2 p^2+(T_3+T_4 )np+C_n n+C_p p}      
 \end{eqnarray}

If we additionally impose charge neutrality ($n=p$), we get the total carrier capture rate due to TAAM and MPE recombination:
 \begin{eqnarray}
R_{tot} &&=-T_2 N^- p^2   -T_3 N^- np -C_p N^- p    \\
  &&= N \frac{(T_1+T_4+C_n/n)(T_2+T_3+C_p/n)} {T_1+T_2+T_3+T_4+(C_n+C_p)/n} n^2
 \end{eqnarray}
 
 Although the TAAM hole capture processes ($T_2$ and $T_3$) start in a negative charge state of the impurity (Ca$_{\rm Ga}^-$), we still use the trap-state wavefunction for the neutral state (Ca$_{\rm Ga}^0$) in our calculations for $T_2$ and $T_3$; adding an electron causes only a minor change in the wavefunction of the in-gap trap state. 

\section{S3. Supercells used in the TAAM calculations}

\begin{figure}[b]
\includegraphics*[width=156mm]{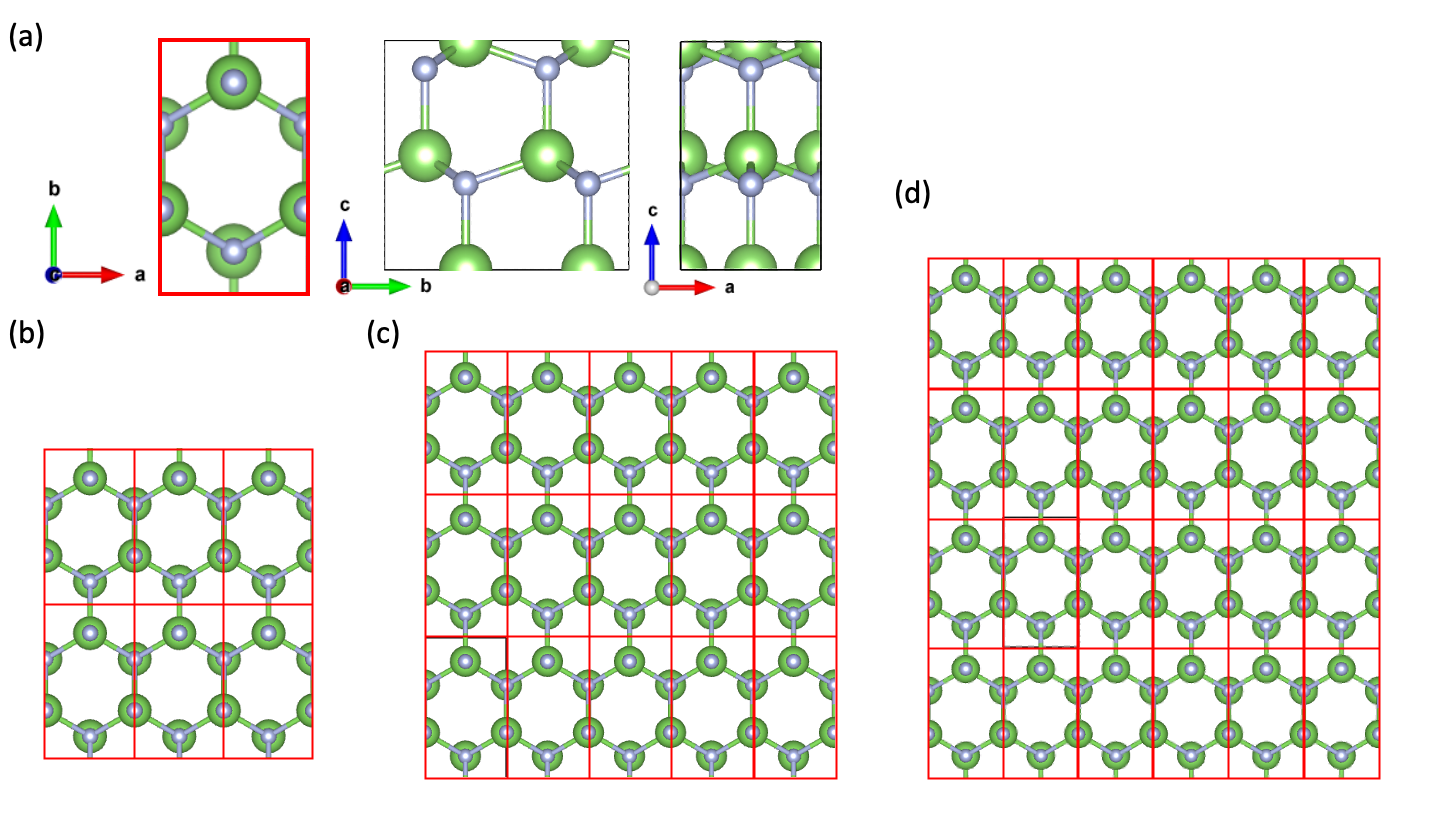}
\caption{\label{fig:S1} 
(a) The 8-atom conventional unit cell of wurtzite GaN.  (b-d) Top view of the 96-atom, 360-atom, and 768-atom supercells.  }
\end{figure}

We construct supercells for wurtzite GaN based on the 8-atom orthorhombic conventional cell depicted in Fig.~\ref{fig:S1}(a). 
The relaxed primitive unit cell of wurtzite GaN in this work has lattice parameters $a$=$b$=3.192 {\AA}, $c$=5.184 {\AA}; the angle between $\bm a$ and $\bm b$ is $60 ^{\circ}$ and the angles between $\bm a$ and $\bm c$ and between $\bm b$ and $\bm c$ are $90 ^{\circ} $.  The orthorhombic conventional cell [Fig.~\ref{fig:S1}(a)] is constructed based on the primitive unit cell of wurtzite GaN by
 \begin{eqnarray}
  \bm a^{conv} &&= \bm a   \nonumber   \\
  \bm b^{conv} &&= \bm a + 2 \bm b   \nonumber   \\
  \bm c^{conv} &&= \bm c      \, .
 \label{eq:conventionalcell}
 \end{eqnarray}
The 96-atom, 360-atom, and 768-atom supercell shown in Figs.~\ref{fig:S1}(b)-(d) are comprised of $3 \times 2 \times 2$, $5 \times 3 \times 3$, and $6 \times 4 \times 4$ multiples of the orthorhombic conventional cell.
These multiples are chosen such that the supercell lattice parameters in three directions are close, ensuring similar separation between impurities in different spatial directions.

 \section{S4. Density-of-states integration method for the $4^{\rm th}$ bulk state in the calculation of $T_i$ coefficients  }
 
\begin{figure}[b]
\includegraphics[width=176mm]{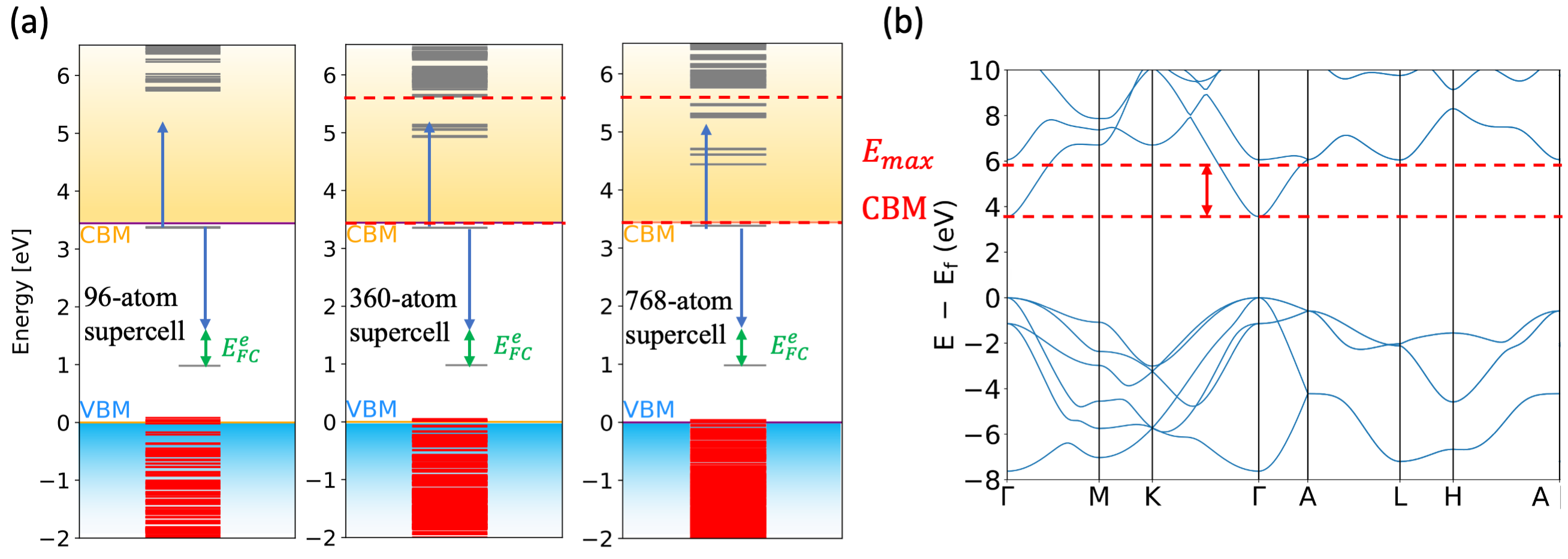}
\caption{\label{fig:S2}  
(a) Diagrams of Kohn-Sham states as calculated with DFT-HSE for a Ca$_{\rm Ga}$ impurity in supercells with 96, 360, and 768 atoms. The blue arrows represent energy conservation and indicate that the 4$^{\rm th}$ state needs to be located $\sim$2.0 eV above the CBM.  
(b) DFT-HSE bulk band structure of GaN, with the CBM and $E_{max}$ indicated.  }
\end{figure}

In this section, we discuss the details of the density-of-state integration method for the $4^{th}$ bulk state in the calculation of $T_i$ coefficients. We take the calculation of $T_1$ as an example. As shown by the band structure in Fig.~\ref{fig:S2}(b) [calculated using density functional theory (DFT) calculation with the HSE functional], the lowest conduction band of GaN exhibits large dispersion.
This renders it very difficult to adequately sample conduction-band (CB) states in calculations with a limited k-point set.
Zone folding in our supercell calculations provides only a sparse sampling of the CB states at the $\Gamma$ point of the supercell, as shown in
Fig.~\ref{fig:S2}(a).
In the $T_1$ process we start with free carriers (electrons) in the conduction band.
Due to the sparseness of CB states in our supercell calculations for $\rm Ca_{Ga}$ in GaN  [Fig.~\ref{fig:S2}(a)], only the state at the CB minimum (CBM) is included as states $\bm 1$ and $\bm 2$ in the calculation of $T_1$. 
 
Energy conservation dictates that the 4$^{\rm th}$ state (in our example) be located $\sim$2.0 eV above the CBM [see the blue arrows in Fig.~\ref{fig:S2}(a)]. 
However, even when including vibrational broadening, due to the sparse sampling of CB states it is unlikely that a relevant KS state will be present in that energy range in the supercell calculations.
Instead of summing over these discrete CB states, we sample the 4$^{\rm th}$ state using integration according to the density of states (DOS) of bulk GaN, denoted by $D(\epsilon_4 )$:
\begin{eqnarray}
T_1 &&= \frac{2\pi}{ \hbar}  \frac{1}{ n^2 }  \sum_{\bm 1 \in c, \bm 2 \in c, \bm 4 \in c}   f_{\bm 1} f_{\bm 2} (1-f_{\bm 4})    | M^1_{\bm 1 \bm 2 \bm t \bm 4} |^2 \, g (\epsilon_{\bm 1}+ \epsilon_{\bm 2} - \epsilon_{\bm t} - \epsilon_{\bm 4} - E_{FC}^e)  ,    \\
&&= \frac{2\pi}{ \hbar}  \frac{1}{ n^2 }  \sum_{\bm 1 \in c, \bm 2 \in c} \int_{CBM}^{\infty}  d\epsilon_{\bm 4} D(\epsilon_{\bm 4}) f_{\bm 1} f_{\bm 2} (1-f_{\bm 4} (\epsilon_{\bm 4} ))     \overline{| M^1_{\bm 1 \bm 2 \bm t \bm 4} |^2}  \, g (\epsilon_{\bm 1}+ \epsilon_{\bm 2} - \epsilon_{\bm t} - \epsilon_{\bm 4} - E_{FC}^e)  ,    \\
&&= \frac{2\pi}{ \hbar}  \frac{1}{ n^2 }  \sum_{\bm 1 \in c, \bm 2 \in c} \int_{CBM}^{ E_{max} }  d\epsilon_{\bm 4} D(\epsilon_{\bm 4}) f_{\bm 1} f_{\bm 2} (1-f_{\bm 4} (\epsilon_{\bm 4}))     \overline{| M^1_{\bm 1 \bm 2 \bm t \bm 4} |^2} \,  g (\epsilon_{\bm 1}+ \epsilon_{\bm 2} - \epsilon_{\bm t} - \epsilon_{\bm 4} - E_{FC}^e)  ,  
 \end{eqnarray}
where in the last line we replaced the upper bound on the integration by $E_{max}$.
For the matrix element $M_{\bm 1 \bm 2 \bm t \bm 4}^1$ we define an averaged value $\overline{| M^1_{\bm 1 \bm 2 \bm t \bm 4} |^2}$:
 \begin{eqnarray}
 \overline{| M^1_{\bm 1 \bm 2 \bm t \bm 4} |^2} = \frac{1}{N_s} \sum_{\epsilon_{CBM} < \epsilon_{\bm 4} < E_{max}} | M^1_{\bm 1 \bm 2 \bm t \bm 4} |^2 \, ,
 \end{eqnarray}
where $N_s$ is the number of 4$^{\rm th}$ states within this energy range between the CBM and $E_{max}$. 

We applied this integration method to the supercell calculations with 360 and 768 atoms (the sampling of CB states in the 96-atom cell is too sparse to give meaningful results). $N_s=12$ for the 360-atom supercell calculation, and $N_s=28$ for the 768-atom supercell calculation. 
 
In order to justify the approximation for the matrix element 
we first inspected the GaN CB states in the energy range $[ \epsilon_{CBM} + \Delta E^{(n/p)} - E_{FC}^{(n/p)} - 2\sigma , \epsilon_{CBM} + \Delta E^{(n/p)} - E_{FC}^{(n/p)} + 2\sigma ]$.
By projecting the wavefunctions onto Ga and N atoms we found that the wavefunction character (embodied in $u_{\bm 4}$) stays almost the same. 
The 360-atom (768-atom) supercell is comprised of a $5 \times 3 \times 3$ ($6 \times 4 \times 4$) multiple of the 8-atom orthorhombic conventional cell of GaN shown in Fig.~\ref{fig:S1}(a) (see Sec.~S3).
In the 360-atom cell the 12 CB states in the relevant energy range are the states arising from zone folding with an effective $k=(\pm \frac{1}{5} \frac{ 2\pi }{a_x}, 0, 0)$, $k=(0, \pm \frac{1}{3} \frac{ 2 \pi }{ a_y}, 0)$, and $k=(0, 0, \pm \frac{1}{3} \frac{2\pi}{a_z})$, with $a_x$, $a_y$, $a_z$ being the lattice parameters of the conventional cell.
``Effective k point'' means that the supercell wavefunction unfolds to the wavefunction with crystal momentum $k$ in the conventional cell. 
Thus, in the matrix element $\bra{\bm 1 \bm 2} \hat{W} \ket{\bm t \bm 4}$ [Eq.~(\ref{eq:ME2})] the integrations  
$ \int  d \bm r_2 u_{\bm 2}^*(\bm r_2) u_{\bm 4} (\bm r_2) e^{i\bm G \cdot \bm r_2}$ (or $ \int  d \bm r_2 u_{\bm 1}^*(\bm r_2) u_{\bm 4} (\bm r_2) e^{i\bm G \cdot \bm r_2}$), with the $4^{th}$ state corresponding to those 12 bulk states, are like a projection of the state $\bm 2$ ($\bm 1$) on a set of plane waves.
This justifies performing the averaging of $|M_{\bm 1 \bm 2 \bm t \bm 4}^1 |^2$ over those $4^{th}$ bulk states.

\section{S5. TAAM coefficients as function of trap-state level position and broadening parameter} 

We investigate the dependence of the TAAM coefficients $T_i$ on the energy of the trap-state level in the band gap.
We start from our calculations for $\rm Ca_{Ga}$ in GaN and use the corresponding wavefunctions for the trap state and all bulk states.
We rigidly shift the energy level of the trap state, which is 0.98 eV for the actual Ca$_{\rm Ga}$ trap, from the valence-band maximum (VBM) to the CBM,
and study the corresponding change of TAAM coefficients.

\begin{figure}
\includegraphics[width=176mm]{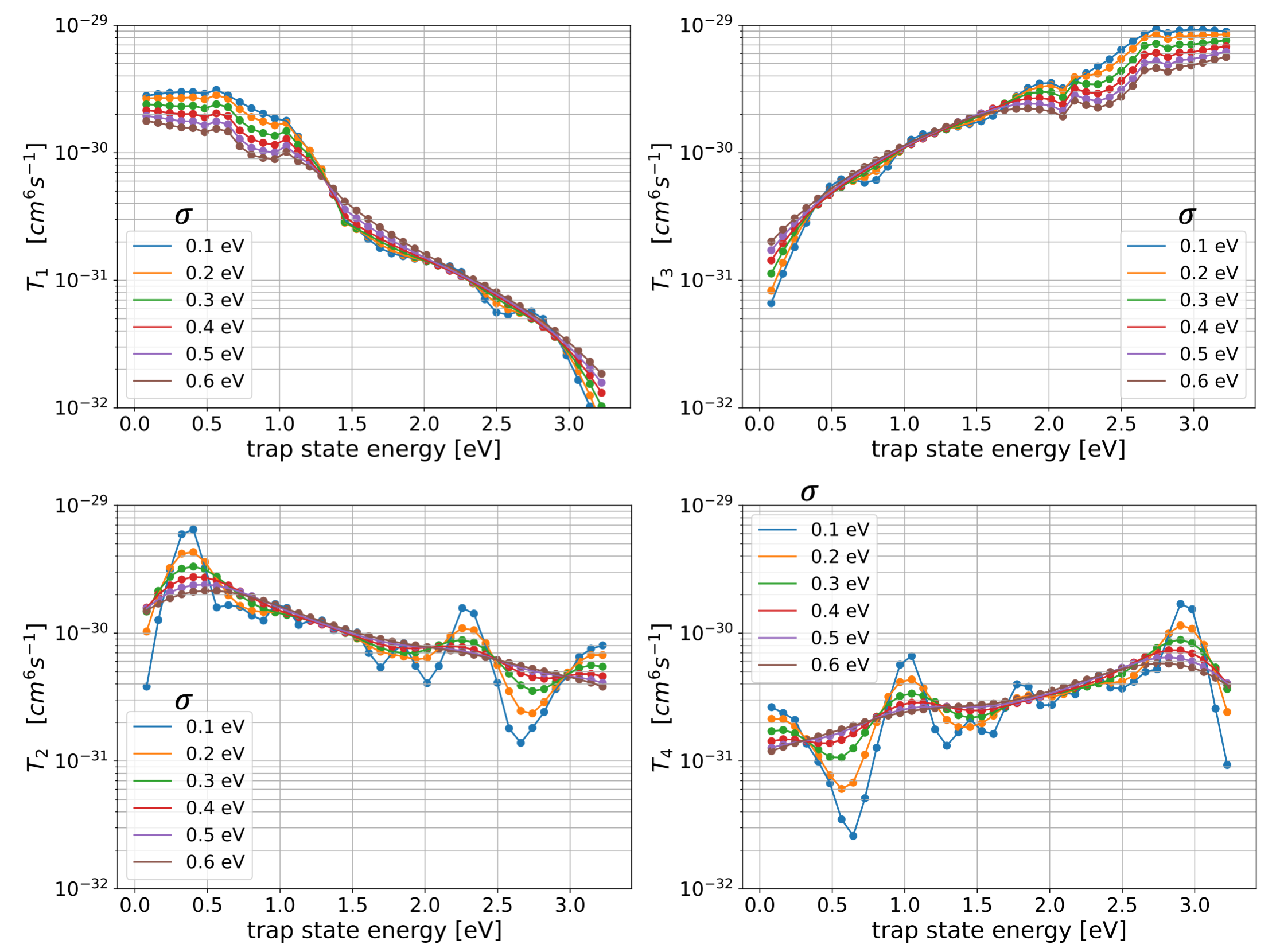} 
\caption{\label{fig:S6} 
Calculated TAAM coefficients for $\rm Ca_{Ga}$ impurities in GaN as a function of the energy of the trap state. Values calculated using different $\sigma$ parameters are denoted by different colors. 
}
\end{figure}

The main result from Fig.~\ref{fig:S6} is that the calculated $T_i$ coefficients depend relatively weakly on the energy of the trap level.  This is in sharp contrast to the dependence observed for capture coefficients for the MPE emission, which depend exponentially on the energy difference between the trap state and the band edge [as is evident, e.g., in Fig.~4(b) of the main text].  The maximum change is by only about two orders of magnitude over the entire range (all the way from a position at the VBM to a position at the CBM).  
The insensitivity to the precise position of the trap level arises from the fact that (unlike MPE) TAAM is not an activated process, and that energy conservation in the TAAM process can easily be satisfied (also facilitated by phonon broadening), since final states are available at essentially any energy and there is no requirement of momentum conservation. We realize of course that for the extreme cases of the trap level near the VBM or the CBM our assumption that the wavefunctions are still the same as for the actual $\rm Ca_{Ga}$ impurity is unlikely to be valid; our main goal was to qualitatively examine the dependence on trap energy.  More quantitatively, the results in Fig.~\ref{fig:S6} allow us to state that uncertainties in the position of the trap level will have a very minor impact on the $T_i$ coefficients: a change in the level by 0.1 eV typically results in a change in $T_i$ by less than 20\%.

The different TAAM processes exhibit different trends as a function of the trap-state energy level. 
The trends are easily explained for $T_1$ and $T_3$, which both involve energy transfer to electrons that are excited to higher-energy states in the CB.  The large dispersion of the GaN CB leads to a significant increase in the DOS of CB states (and hence more final states being available) at higher energies.
$T_1$, which involves electron capture into the trap level, increases as the trap-state energy goes down; indeed, a larger transition energy allows excitation of an electron into higher-energy states in the CB, where the DOS is larger.
A similar trend ({\it mutatis mutandis}) is observed for the hole capture process $T_3$: larger transition energies and hence larger values of $T_3$ are observed for trap-state energies higher in the gap.

For $T_2$ and $T_4$ the trend is different.
Apparently in this case the larger density of states at energies farther from the VBM plays less of a role; indeed, the increase in the VB DOS is less pronounced than the increase in the CB DOS.
More importantly, the variation in $T_2$ and $T_4$ appears to be governed by a change in the matrix elements, which decrease as the transition energy increases.

In Fig.~\ref{fig:S6} we also analyze the dependence of the results on the value of the  broadening parameter $\sigma$ in the Gaussion used to describe the electron-phonon interaction [see Eq.~(10) of the main text]. 
The broadening parameter $\sigma_e$ ($\sigma_h$) for the electron (hole) capture process are estimated based on the 1D configuration coordinate diagram \cite{jia2020design, alkauskas2016tutorial}: 
 \begin{eqnarray}
\sigma_e &&=  \hbar \Omega_e \sqrt{S^e } \sqrt{\cot \left(\frac{\hbar\Omega_e}{2k_B T} \right)}  \\
\sigma_h &&=  \hbar \Omega_g \sqrt{S^h } \sqrt{\cot \left(\frac{\hbar\Omega_g}{2k_B T} \right)} \, .
 \label{eq:sigma}
 \end{eqnarray}
$\Omega_e$ and $\Omega_g$ are the effective phonon frequency of the $-$1 charge state and the neutral charge state, respectively, in a 1D harmonic approximation. 

As noted in the main text, a value $\sigma$=0.22 eV provides a good fit.
However, as seen in Fig.~\ref{fig:S6}, other values (ranging from 0.1 to 0.6 eV) 
lead to only small variations in the $T_i$ values (by a factor of 2 or less).
(The oscillations in $T_2$ and $T_4$ with smaller $\sigma$ arise from imperfect sampling to the 4th state in the VB, for which we did not apply the DOS integration method.)
It is reassuring that the $T_i$ coefficients are not sensitive to the precise magnitude of the broadening.

\bibliography{SM_bibliography}